\documentclass[10pt,conference]{IEEEtran}
\IEEEoverridecommandlockouts
\usepackage{cite}
\usepackage{amsmath,amssymb,amsfonts}
\usepackage{graphicx}
\usepackage{textcomp}
\usepackage{xcolor}
\usepackage[T1]{fontenc}
\usepackage{listings}
\usepackage{balance}
\usepackage{longtable}

\usepackage[ruled,linesnumbered]{algorithm2e}
\usepackage{wrapfig}
\usepackage{makecell}
\usepackage{subfigure}
\usepackage{multirow}
\usepackage{color}
\usepackage{soul}
\usepackage{booktabs}
\usepackage{moreverb}
\usepackage{enumitem}
\usepackage{graphicx}
\usepackage{epstopdf}
\usepackage{subfigure}
\usepackage{miniplot}
\usepackage{afterpage}
\usepackage[most]{tcolorbox}
\usepackage{float}
\usepackage{hyperref}
\usepackage{array}
\usepackage{xspace}
\usepackage{mathtools}

\SetKwInput{kwInit}{Init}

\def\BibTeX{{\rm B\kern-.05em{\sc i\kern-.025em b}\kern-.08em
    T\kern-.1667em\lower.7ex\hbox{E}\kern-.125emX}}
\begin{document}

\newcommand{\tool}{\textsc{Kpc}\xspace}
\newcommand{\xx}[1]{\textcolor{orange}{\textbf{XX:}#1}}
\newcommand{\tabincell}[2]{\begin{tabular}{@{}#1@{}}#2\end{tabular}}
\newcommand{\zc}[1]{\textcolor{red}{\textbf{ZC:}#1}}

\makeatletter
\newcommand{\linebreakand}{%
  \end{@IEEEauthorhalign}
  \hfill\mbox{}\par
  \mbox{}\hfill\begin{@IEEEauthorhalign}
}
\makeatother

\title{From Misuse to Mastery: Enhancing Code Generation with Knowledge-Driven AI Chaining}

\author{\IEEEauthorblockN{1\textsuperscript{st} Xiaoxue Ren}
\IEEEauthorblockA{\textit{School of Software Technology} \\
\textit{Zhejiang University}\\
Hangzhou, China \\
xxren@zju.edu.cn}
\and
\IEEEauthorblockN{2\textsuperscript{nd} Xinyuan Ye}
\IEEEauthorblockA{\textit{School of Computing} \\
\textit{Australian National University}\\
Canberra, Australia \\
xinyuan.ye@anu.edu.au}
\and
\IEEEauthorblockN{3\textsuperscript{rd} Dehai Zhao}
\IEEEauthorblockA{\textit{CSIRO's Data61} \\
%\textit{Australian National University}\\
Sydney, Australia \\
dehai.zhao@data61.csiro.au}
\linebreakand 
\IEEEauthorblockN{4\textsuperscript{th} Zhenchang Xing}
\IEEEauthorblockA{\textit{CSIRO's Data61} \\
\textit{\& Australian National University}\\
Sydney, Australia \\
zhenchang.xing@data61.csiro.au}
\and
\IEEEauthorblockN{5\textsuperscript{th} Xiaohu Yang}
\IEEEauthorblockA{\textit{College of  Computer Science and Technology} \\
\textit{Zhejiang University}\\
Hangzhou, China \\
yangxh@zju.edu.cn}
% \and
% \and
% \and
}

\maketitle

\begin{abstract}
Large Language Models (LLMs) have shown promising results in automatic code generation by improving coding efficiency to a certain extent.
However, generating high-quality and reliable code remains a formidable task because of LLMs' lack of good programming practice, especially in exception handling.
In this paper, we first conduct an empirical study and summarize three crucial challenges of LLMs in exception handling, i.e., incomplete exception handling, incorrect exception handling and abuse of try-catch. 
We then try prompts with different granularities to address such challenges, finding fine-grained knowledge-driven prompts works best.
Based on our empirical study, we propose a novel \underline{K}nowledge-driven \underline{P}rompt \underline{C}haining-based code generation approach, name \tool, which decomposes code generation into an AI chain with iterative check-rewrite steps and chains fine-grained knowledge-driven prompts to assist LLMs in considering exception-handling specifications.
We evaluate our \tool-based approach with 3,079 code generation tasks extracted from the Java official API documentation.
Extensive experimental results demonstrate that the \tool-based approach has considerable potential to ameliorate the quality of code generated by LLMs. 
It achieves this through proficiently managing exceptions and obtaining remarkable enhancements of 109.86\% and 578.57\% with static evaluation methods, as well as a reduction of 18 runtime bugs in the sampled dataset with dynamic validation.
\end{abstract}

\begin{IEEEkeywords}
Large Language Model, Code Generation, Knowledge-driven Prompt, API Misuse
\end{IEEEkeywords}

\section{Introduction}
% LLMs in code generate
Large Language Models (LLMs) have gained significant attention in the field of natural language processing (NLP) for their ability to generate coherent and contextually relevant text~\cite{arora2022ask, schick2022peer, yang2022re3, wu2022ai, creswell2022selection, kazemi2022lambada}.
Recently, there has been growing interest in using LLMs for code generation. 
LLMs, such as Codex~\cite{chen2021evaluating}, AlphaCode~\cite{li2022competition}, CODEGEN~\cite{nijkamp2022codegen}, and INCODER~\cite{fried2022incoder}, use sophisticated algorithms to generate code based on natural language input, enabling developers to considerably automate the coding process.
The use of LLMs in code generation holds great promise for improving software development processes by reducing the time and effort required for coding tasks and increasing developers' productivity~\cite{vaithilingam2022expectation,dong2023self}.

% limitations of LLMs in code generation
Nevertheless, generating high-quality and reliable code remains a formidable task~\cite{barke2023grounded,gozalo2023chatgpt,mastropaolo2023robustness,dong2023self}.
One significant limitation of LLMs in code generation is the lack of good programming practice.
LLMs are based on statistical patterns and lack the ability to understand the underlying logic and structure of programming languages. 
Therefore, they may generate code that is syntactically correct but not semantically correct, resulting in poor programming practices.
Particularly, LLMs have challenges in generating reliable, maintainable and robust code in terms of exception handling, which has been proven to be essential in software development~\cite{leinonen2023using,barbosa2018global,de2018studying}.

\begin{figure}[t]
  \centering
\includegraphics[width=0.43\textwidth]{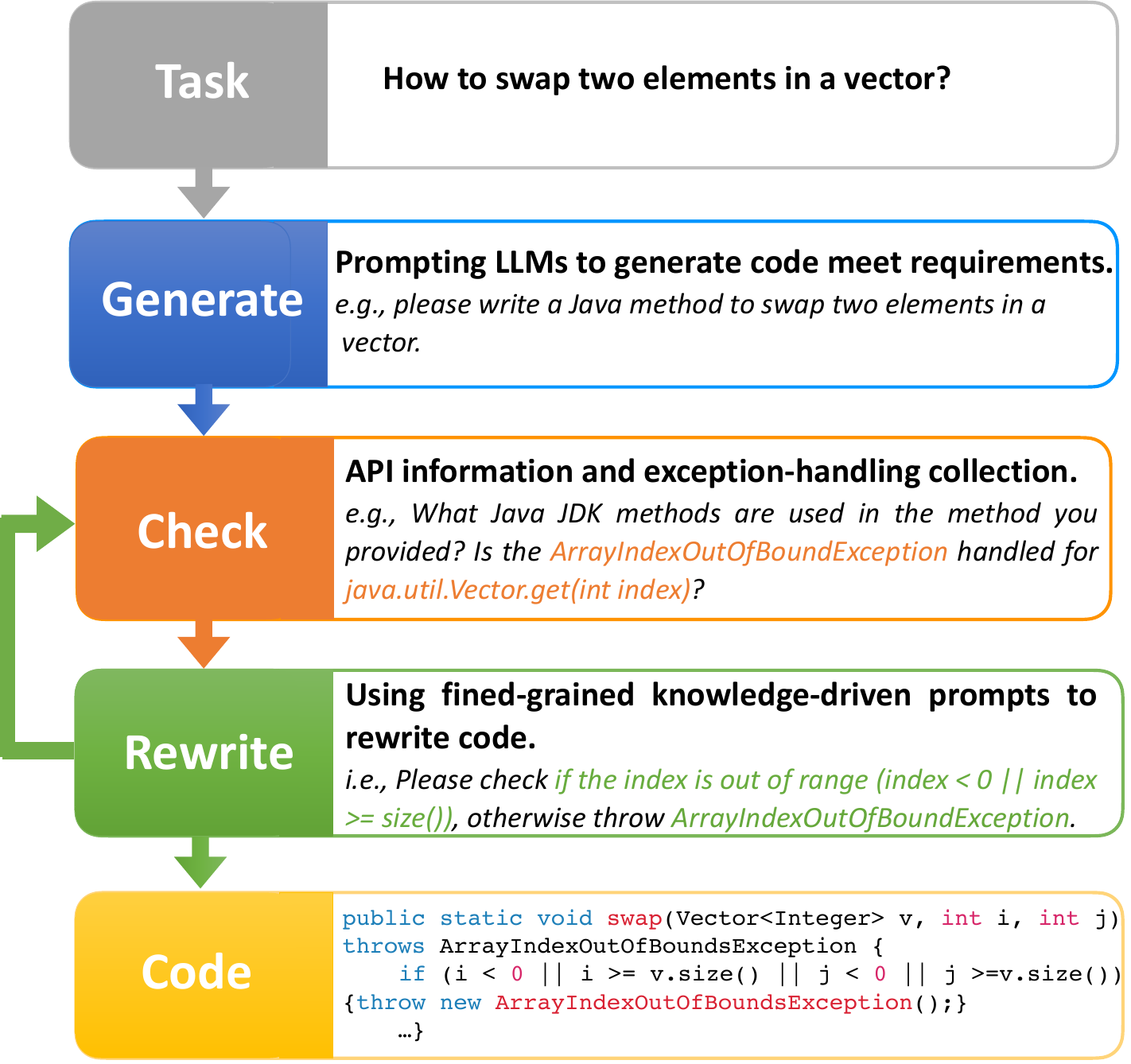}
\vspace{-3mm}
  \caption{High-level Overview of \tool-based Code Generation.}
  \label{fig:overall}
  
\end{figure}
% challenges we summarized in the empirical study
Three crucial exception-handling challenges of code generated by LLMs have been summarized by our empirical study in Section~\ref{sec:motivate_challenges}, including \emph{\textbf{incomplete exception handling}}, \emph{\textbf{incorrect exception handling}}, and \emph{\textbf{abuse of \textsf{\small{try-catch}}}}. 
For example, code in Figure~\ref{fig:motivation} (A) demonstrates an incomplete exception handling example, as both \textsf{\small{java.util.Vector.get(int index)}} and \textsf{\small{java.util.Vector.set(int index, E element)}} may encounter exceptions.
Code in Figure~\ref{fig:motivation} (B) shows an incorrect exception handling example, in which \textsf{\small{ArrayIndexOutOfBoundsException}} should be handled instead of \textsf{\small{IndexOutOfBoundsException}}.
Moreover, the usage of \textsf{\small{try-catch}} statements in Figure~\ref{fig:motivation} (C) is not considered to be the best practice for exception handling.
Such challenges will result in serious consequences, including software crashes and reliability and security issues~\cite{sena2016understanding,zhang2020learning}.

% different prompts in dealing with exception handling issues.
\begin{comment}
To address these challenges, prompting exception handling can be utilized, where we prompt LLMs with additional information to help them understand the expected behavior of the code and how to handle exceptions that may occur.
However, it is vital to try different prompts when prompting exception handling for code generation because there may be multiple ways to handle exceptions depending on the specific scenario. 
By trying different prompts in the empirical study (see Section~\ref{sec:motivate_prompts}), we can explore various possible scenarios and edge cases, and enable the large model to iterate and address potential exceptions autonomously.
For example, part $B$, $c_1$, and $c_2$ in Fig.\ref{fig:motivation} show results with different prompts.
From empirical study, we observe using fine-grained knowledge-driven prompts is most effective in preventing incomplete or incorrect exception handling, as well as the abuse of \textsf{\small{try-catch}} statement.
\end{comment}

In order to obtain ideal results from LLMs, prompt engineering~\cite{gu2021ppt, liu2022p, chen2022knowprompt, liao2022ptau} is one of the most effective solutions and has been widely studied.
This technique focuses on designing proper prompts that guide LLMs to take desired actions, with the goal of improving the quality of output.
In the context of software engineering, prompt engineering can help LLMs complete a series of development tasks with improved user experiences, reduced errors and support costs, and increased user adoption and satisfaction~\cite{chen2022knowprompt,nashid2023retrieval}.
In our work, a well-designed prompt can make LLMs understand the expected behavior of coding and find the best solution to handle exceptions in the code.
With different prompts in the empirical study (see Section~\ref{sec:motivate_prompts}), we can explore various possible scenarios and edge cases, and enable LLMs to iterate and address potential exceptions autonomously.
Figure~\ref{fig:motivation} (B), (C), and (D) show responses of LLMs with the different granularities of prompts, where we find fine-grained knowledge-driven prompts are most effective in preventing incomplete or incorrect exception handling, as well as the abuse of \textsf{\small{try-catch}} statement.

% our kpc-based approach
In this paper, we propose a novel \textbf{K}nowledge-driven \textbf{P}rompt \textbf{C}haining-based code generation approach, named \tool, which utilizes fine-grained exception-handling knowledge extracted from API documentation to assist LLMs in code generation.
This approach follows the divide-and-conquer strategy and iterative coding practice, which improves the generated code from misuse to mastery by formulating a chain of modular prompts for LLMs.
To this end, we first construct an API knowledge base from official API documentation~\cite{javadoc}. 
Then, based on the original code generated by LLMs, we chain knowledge-driven prompts to check whether there exists any unhandled exception in the generated code. 
Whenever such exceptions are identified, we rewrite the code with fine-grained exception-handling knowledge prompts until all exceptions are properly handled, as shown in Figure~\ref{fig:overall}.

The API knowledge base constructed for our \tool-based code generation approach includes 19,057 exceptions with corresponding conditions, covering 11,477 Java SDK \& JDK APIs, which is available in our replication package~\cite{KPCgithub}.
We conduct experiments to evaluate the efficiency and effectiveness of our \tool-based code generation approach for exception handling. 
Our experiments are based on 3,079 Java coding tasks obtained from Java SDK \& JDK API functional descriptions~\cite{javadoc}.
To evaluate the efficiency, we analyze the distributions of unhandled exceptions and checking-rewriting loops.
We find that our \tool-based approach can efficiently handle exceptions for the majority of coding tasks using a small number of iterative checking-rewriting loops. 
Specifically, 90.35\% of the generated code can be effectively handled within ten loops when dealing with exceptions.
To evaluate the effectiveness, we use three evaluation methods, including both static and dynamic validations. 
For static validations, on one hand, we leverage the ability of LLMs to automatically evaluate the exception-handling practice of the generated code. 
On the other, we manually review a subset of the tasks. 
For dynamic validation, we employ EvoSuite~\cite{fraser2011evosuite} to generate test cases for detecting runtime bugs.
The results of all three evaluation methods show that our \tool-based approach outperforms the state-of-the-art code generator (i.e., ChatGPT) in exception handling. 
Specifically, we observed improvements of 109.86\% and 578.57\% in the two static evaluation methods, and a reduction of 18 runtime bugs in the dynamic validation.
In addition to these evaluations, we also conduct a user study to determine how our \tool-based approach can help developers in practice. 
Results show that our approach can effectively remind developers of exception-handling specifications, leading to a high level of correctness (75.00\%) in handling potential exceptions.

In summary, we make the following contributions:
\begin{itemize}
    \item We are the first to propose the fine-grained knowledge-driven prompt chaining approach (i.e., \tool) for LLMs in code generation tasks.
    \item We conduct an empirical study to investigate the challenges of LLMs in exception handling, and design the most appropriate prompts to address these challenges.
    \item Experimental results demonstrate that our \tool-based approach is highly efficient and effective in handling exceptions. And our user study also shows \tool can assist developers in effectively handling exceptions in practice.
    \item We open source our replication package~\cite{KPCgithub}, including the dataset, the source code of \tool, and experimental results, for follow-up works.
\end{itemize}

\section{An Empirical Study on Exception Handing for LLMs Code Generation}\label{sec:empirical}

Despite the great success of LLMs in the software engineering field, there still exist some issues that have not been well studied in the literature, leading to unexpected errors when solving tasks such as code generation in this work.
Therefore, we conduct an empirical study to facilitate the understanding of knowledge-driven prompts chaining for LLMs in code generation by answering the following two research questions.
% In the study, we first collect 183 Java code generation tasks from tutorialspont~\cite{Tutorialspoint}, which can be accessed at our replication package~\footnote{\url{https://public}}.
% Then, we rephrase the task as inputs of ChatGPT to generate Java code snippets.
\begin{itemize}[leftmargin=*, topsep=0pt]
\item \textbf{RQ1: What challenges do LLMs face in handling exceptions when generating code?} By analyzing the generated code from LLMs, we manually summarize the challenges related to exception handling. 
\item \textbf{RQ2: How to help LLMs address the challenges effectively?} 
After summarizing the challenges, we leverage prompt engineering to address them by trying different prompts related to exception handling. 
\end{itemize}

\subsection{Challenges of LLMs in Exception Handling (RQ1)}~\label{sec:motivate_challenges}
To explore the challenges that LLMs encounter in handling exceptions, we first collect 92 Java code generation tasks from Tutorialspoint~\cite{Tutorialspoint}, which may have potential exceptions.
We then rephrase (see Section~\ref{sec:generating}) the tasks as inputs of ChatGPT to generate Java code.
Afterward, two of the authors collaborate to review the generated code manually with the goal of identifying the challenges that LLMs might encounter when handling exceptions in code generation tasks.
Both authors have more than five years of experience in Java development and a thorough understanding of official Java API documentation.
We summarize the challenges into three crucial aspects, which are elaborated as follows.

\subsubsection{Incomplete exception handling}
%definition
Exceptions are an essential part of programming that signal the occurrence of an error during program execution. It's common for a single code snippet to encounter multiple exceptions. However, failing to handle all of these exceptions can have severe consequences, such as crashes, data corruption, and security vulnerabilities.
Unfortunately, results of the empirical study show that the incomplete exception handling phenomenon is very common in the code generated by ChatGPT, since there are 88.04\% (81 out of 92) of the coding tasks have such issues.
In fact, such a fault is very likely to be avoided because official documentation clearly describes the exception-handling specifications for each API.
For example, the code in Figure~\ref{fig:motivation} (A) involves the incomplete exception-handling issues for the two APIs, \textsf{\small{java.util.Vector.get(int index)}} and \textsf{\small{java.util.Vector.set(int index, E element)}}, and we can find the exception-handling specification \emph{``Throws: ArrayIndexOutOfBoundsException - if the index is out of range (index < 0 || index >= size())''} in Java API documentation~\cite{specification4get,specification4set}.

\begin{comment}
According to the empirical study, it is common that incomplete exception handling issues happen in code generated by ChatGPT, namely 88.04\% (81 out of 92 coding tasks) generated code encounter such issues. 

%example
For example, Figure~\ref{fig:motivation} (A) shows an intuitive code example generated by ChatGPT, which with incomplete exception-handling issues.
Specifically, both the Java APIs (i.e., \textsf{\small{java.util.Vector.get(int index)}} and \textsf{\small{java.util.Vector.set(int index, E element)}}) used in the generated code have the exception-handling specification:\emph{``Throws: ArrayIndexOutOfBoundsException - if the index is out of range (index < 0 || index >= size())''}, according to Java official documentation.
Hence, \textsf{\small{ArrayIndexOutOfBoundsException}} should be handled to avoid unexpected errors in the generated code.
\end{comment}

% \textcolor{blue}{Not sure if it is necessary to write conclusion here}
% To conclude, it is important for developers to carefully plan and implement exception handling in their code to ensure that errors are caught and properly handled, and to minimize the impact of unexpected exceptions on the system as a whole. 

\subsubsection{Incorrect exception handling}
\label{sec:incorrect_exception_handling}
\begin{comment}
This issue occurs when an exception is handled incorrectly in the generated code, such as catching the wrong type of exceptions.
Besides crashes, incorrect exception handling can also make it more difficult to debug problems in the program. 
This is because the code may fail in unexpected ways, and it may be difficult to determine the root cause of the problem.
According to the empirical study, the incorrect exception handling is also rather serious, where 4.35\% (4 out of 92) code directly generated by ChatGPT has handled the incorrect exceptions.
\end{comment}
This issue occurs when an exception is handled incorrectly in the code, such as catching the wrong type of exceptions.
According to the statistical results of our empirical study, 4.35\% (4 out of 92) of the code examples generated by ChatGPT contain such issues.
Although these fault examples account for a small part of the overall coding tasks, we cannot ignore this issue, especially for those code examples that are able to run without any error, which would be very hard to localize the root cause of the problem.
Figure~\ref{fig:motivation} (B) shows an example of incorrect exception handling, which catches \textsf{\small{IndexOutOfBoundsException}}.
Referring to Java API documentation~\cite{specification4get,specification4set}, it is recommended to use \textsf{\small{ArrayIndexOutOfBoundsException}} instead of \textsf{\small{IndexOutOfBoundsException}} in this scenario, as \textsf{\small{ArrayIndexOutOfBoundsException}} is a subclass of \textsf{\small{IndexOutOfBoundsException}} and is specifically designed for cases where an array or vector index is out of bounds.

% Overall, it is important for developers to carefully plan and test their exception-handling code to ensure that it properly handles the correct possible scenarios and leads to correct behavior in the event of an exception.

\subsubsection{Abuse of try-catch}
\label{sec:abuse_of_try_catch}
It refers to situations where \textsf{\small{try-catch}} statements are used excessively or inappropriately, e.g., using them to handle non-exceptional situations or to handle exceptions that should be handled by a specific piece of code. 
We observe from the empirical study that ChatGPT attempts to use \textsf{\small{try-catch}} statements to solve 5.43\% (5 out of 92) of the coding tasks, which constitutes a small percentage of the whole dataset, but we can not take this issue lightly, as this practice will reduce the readability and maintainability of the code.
% Usually, a \textsf{\small{try-catch}} statement makes the code physically executable but not logically reasonable, and a more effective solution is desired to handle the exception.
Typically, a \textsf{\small{try-catch}} statement enables code to run without errors, but it may not always lead to logically sound code. 
Hence, it's often necessary to find a more effective solution to handle exceptions.
For example, the code in Figure~\ref{fig:motivation} (C) implements the function of throwing exceptions if there exists any error with the input, but it lacks the ability to determine the explicit conditions to throw exceptions.
Instead, using an \textsf{\small{if-condition}} checking mechanism implemented in Figure~\ref{fig:motivation} (D), is generally considered to be a better practice in the software development process.

\begin{comment}
This practice can result in code that is difficult to read or maintain, as well as negatively impact the performance of the program.
According to the empirical study, abuse of \textsf{\small{try-catch}} is also serious in code generated by ChatGPT, where 5.43\% (5 out of 92) generated code has such issues.
For example, in the context of generated programs, the use of \textsf{\small{try-catch}} blocks in Figure~\ref{fig:motivation} (A), (B) and (c1) is considered inappropriate. 
This is because these blocks are not designed to handle non-exceptional situations or to replace the functionality of calling code. 
Instead, using an \textsf{\small{if-condition}} checking mechanism, such as that implemented in Figure~\ref{fig:motivation} (c2), is generally considered to be a better practice for handling such exceptions.
\end{comment}

% Overall, it is important to use \textsf{\small{try-catch}} statements judiciously and only for their intended purpose of handling exceptional situations. 
% Excessive or inappropriate use can result in code that is difficult to maintain and may negatively impact program performance.
\vspace{-1.5mm}
\begin{tcolorbox}[breakable,width=\linewidth-2pt,boxrule=0pt,top=2pt, bottom=2pt, left=2pt,right=2pt, colback=gray!20,colframe=gray!20]
\textbf{Answer to RQ1:} Code generated by LLMs mainly encounters three challenges, including \emph{incomplete exception-handling}, \emph{incorrect exception-handling} and \emph{abuse of \textsf{\small{try-catch}} statements}.
% , among which the incomplete exception handling is the most serious.
\end{tcolorbox}

\begin{figure}[t]
  \centering
  \includegraphics[width=0.43\textwidth]{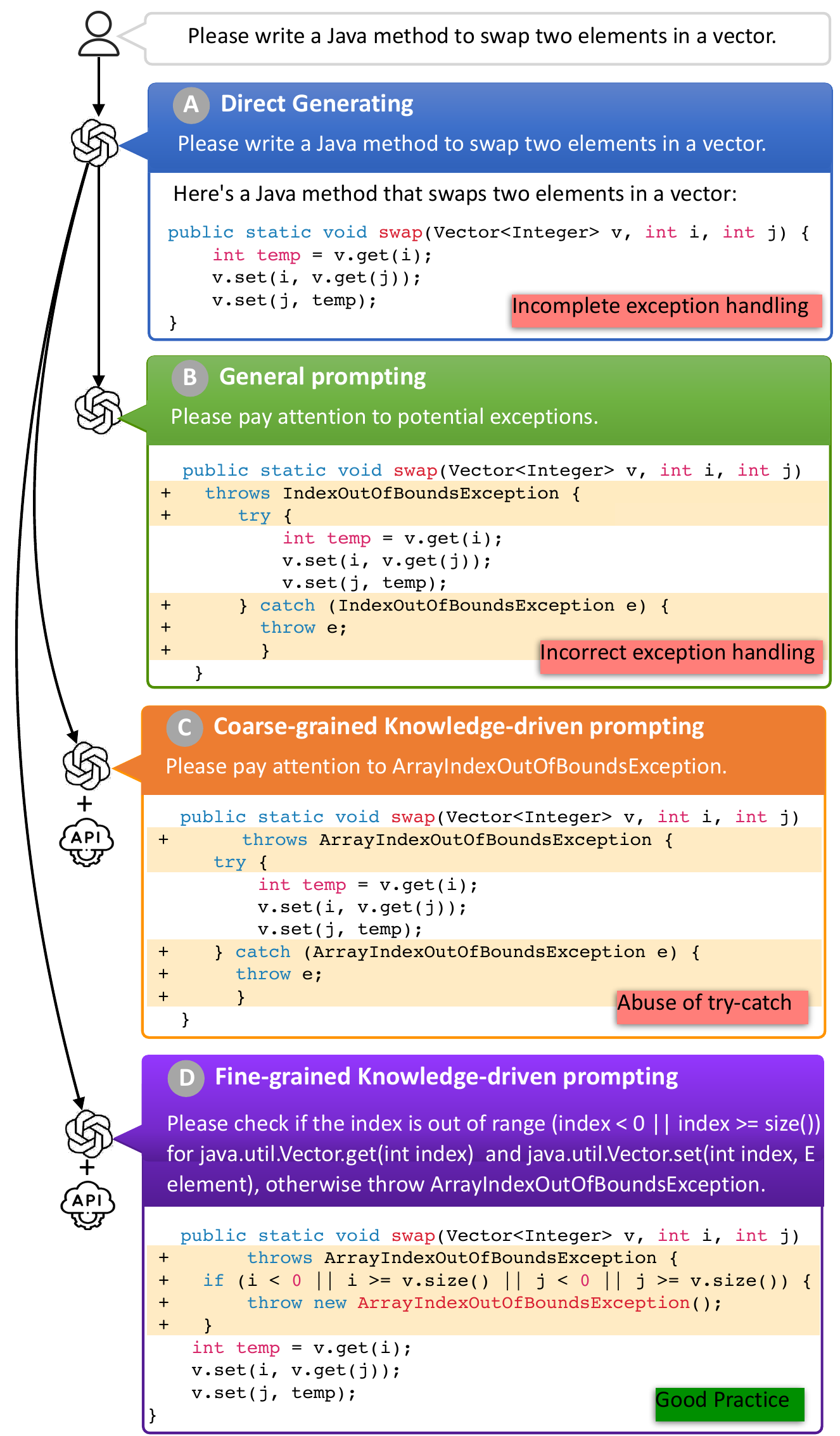}
  \vspace{-3mm}
  \caption{A walkthrough example demonstrating the difference between direct generating (A), general prompting (B), and both coarse-grained (C) and fine-grained (D) knowledge-driven promoting, using the coding task from Tutorialspoint.}
  \label{fig:motivation}
\end{figure}

\subsection{Prompt Engineering for Exception Handling (RQ2)}\label{sec:motivate_prompts}
\begin{comment}
As mentioned earlier, the code generated by LLMs faces many challenges in exception handling.
By providing prompts that focus on exception handling, we can guide LLMs to generate code that is better equipped to handle a variety of error conditions.
However, various outcomes are evident when using different prompts.
To determine the most suitable \xx{exception-handling prompts} for LLMs, our empirical study endeavors involved experimenting with various granularity of prompts, including general prompts, coarse-grained and fine-grained prompts.
\end{comment}

As shown in RQ1, LLMs have great potential to solve software engineering tasks such as code generation, but the three challenges are still significant barriers to exception handling.
It is believed that prompt engineering is the key to unlocking the magic of LLMs to generate high-quality solutions.
However, prompt engineering involves much more than just developing prompts.
It requires a deliberate and systematic approach to designing and refining prompts and the underlying data structures that control LLMs.
%The prompts in code generation tasks is analogous to the requirement clarification in software development life cycle.
In this RQ, we apply a three-granularity (i.e. general, coarse-grained and fine-grained) prompt design strategy, which adds exception-handling relevant information to prompts gradually, to investigate the impact of prompts to exception handling in LLM-based code generation tasks.

\subsubsection{General prompting}
%definition
\begin{comment}
It refers to a type of prompt that is broad in nature and commonly used to attract LLMs' attention to potential exceptions. 
The prompt \emph{``Please pay attention to potential exceptions.''} in Figure~\ref{fig:motivation} (B) is an example of general prompts. 

%strength of general prompting
Compared to directly generated code, general prompting is able to activate the ability of LLMs to handle exceptions with minimal effort. This, in turn, reduces incomplete exception handling from 81 to 35 cases (as shown in the General Prompting column of Table ~\ref{tab:statistics_of_empirical_study}).
Specifically, the code in Figure~\ref{fig:motivation} (B) demonstrates a higher level of awareness regarding exception handling when compared to the code generated directly by ChatGPT (i.e., the code in Figure~\ref{fig:motivation} (A)).
\end{comment}

The most straightforward way to make LLMs pay attention to the exceptions in code is asking them to do so directly, which is named as general prompting in our work.
This method only uses generic warnings to raise concerns about exception to LLMs without providing exception types and conditions.
Figure~\ref{fig:motivation} (B) shows an example of general prompting, in which we give the prompt \emph{``Please pay attention to potential exceptions.''} to ChatGPT.
By reviewing the code generated in the last round, ChatGPT realizes there exists potential exceptions involving ``out of bounds'' and solves this problem by catching \textsf{\small{IndexOutOfBoundsException}}.
Although this method tries to solve problems with minimal effort, it can solve a part of the exception-handling issues.
According to the experiment results in Table~\ref{tab:statistics_of_empirical_study}, general prompting helps ChatGPT to reduce incomplete exception handling examples from 81 to 35.
Nevertheless, this solution may introduce other problems, such as incorrect exception handling (Section~\ref{sec:incorrect_exception_handling}) and abuse of try-catch (Section~\ref{sec:abuse_of_try_catch}) because the general prompting fails to provide sufficient information about the types of exceptions in the code and appropriate measures to handle them.
As presented in Table~\ref{tab:statistics_of_empirical_study}, comparing general prompting with direct generating, the number of incorrect exception handling increases from 4 to 23, and the number of abusing try-catch goes up from 5 to 20.
Without comprehensive exception-handling specifications, LLMs create low-quality solutions to the coding tasks, which go against the best practices in the software development process.

\subsubsection{Coarse-grained prompting}
\begin{comment}
It refers to a type of prompt that provides specific exceptions that should be considered. 
It is more focused than a general prompt but less specific than a fine-grained prompt.
The prompt \emph{``Please pay attention to ArrayIndexOutOfBoundsException.''} in Figure~\ref{fig:motivation} (c1) is an example of coarse-grained prompts.

% strength
Since coarse-grained prompts provide specific exceptions for handling, there is less chance of handling incorrect exceptions, reducing the chances of bugs and errors.
Table~\ref{tab:statistics_of_empirical_study} shows that the implementation of coarse-grained prompting led to a significant reduction of incomplete exception handling issues by 95.06\% (from 81 to 4).
Additionally, the number of incorrect exception handling was completely eliminated due to the effectiveness of the prompt system.
For example, by providing specific exceptions for handling (e.g., Figure~\ref{fig:motivation} (c1)), the refined code can accurately handle exceptions like \textsf{\small{ArrayIndexOutOfBoundsException}}.
\end{comment}

Compared with general prompting which provides zero exception-handling information, coarse-grained prompting offers a part of such information to the input prompts of ChatGPT.
Specifically, this strategy applies exception-handling specifications from API documentation and indicates what exception ChatGPT should pay attention to exactly.
As shown in Figure~\ref{fig:motivation} (C), the prompt is updated to \emph{``Please pay attention to ArrayIndexOutOfBoundsException.''} from the naive version.
ChatGPT understands clearly that it should focus on the given exception and revises the code by catching \textsf{\small{ArrayIndexOutOfBoundsException}} with a \textsf{\small{try-catch}} block.
Based on the statistical results in Table~\ref{tab:statistics_of_empirical_study}, coarse-grained prompting achieves a significant reduction of incomplete exception-handling issues by 95.06\% (from 81 to 4).
Moreover, the number of incorrect exception handling is completely eliminated.
Theoretically, if ChatGPT is provided with explicit exception information, there will be no chance to exist incomplete or incorrect exception handling.  
We investigate the reason why there still exist four incomplete exception-handling examples and find they use APIs whose exception-handling specifications are incomplete in the official documentation.
For example of \textsf{\small{java.lang.String.split(String regex)}}~\cite{specification4split}, neither of the object string nor parameter regex can be \textsf{\small{null}}, otherwise \textsf{\small{NullPointerException}} will happen. 
Therefore, the prompts are not given such information, and the corresponding exceptions cannot be handled.
We also observe some side effects of this strategy from Table~\ref{tab:statistics_of_empirical_study}, in which the number of abusing \textsf{\small{try-catch}} statements by coarse-grained prompting increases from 5 to 41 compared with direct generating.
The main reason lies in the lack of condition information of the exceptions, so ChatGPT is unable to handle them in appropriate manners. 

\begin{comment}
%weakness
However, coarse-grained prompts do not provide fine-grained control over the generated code.
This can also possibly lead to the phenomenon of abuse of \textsf{\small{try-catch}} statements.
From the Coarse-grained column of Table~\ref{tab:statistics_of_empirical_study}, the number of abusing \textsf{\small{try-catch}} statements is increased from 5 to 41.
Despite being refined with coarse-grained prompts, the resulting code still relied on a \textsf{\small{try-catch}} statement, highlighting the need for more precise prompts to avoid such issues.
\end{comment}

% To conclude, coarse-grained prompts can provide LLMs with specific exceptions to handle, but still cannot be sufficient for more complex scenarios or for ensuring compliance with established industry best practices.

\subsubsection{Fine-grained prompting}
From the previous prompting strategies, we find that the more detailed information given to prompts, the higher quality can ChatGPT solve the problems.
Therefore, in the final edition of prompt engineering, we introduce all the exception-handling information obtained from official Java API documentation and propose fine-grained prompting strategy, which contains not only the exception types, but also corresponding conditions to trigger the exceptions.
As shown in Figure~\ref{fig:motivation} (D), the prompt is formulated as \emph{``Please check if the index is out of range (index $< 0 $ $||$ index $>=$ size()) for  java.util.Vector.get(int index) and java.util.Vector.set(int index, E element), otherwise throw ArrayIndexOutOfBoundsException.''}.
ChatGPT leverages the key information from the prompt and provides the code solution by adding an \textsf{\small{if-condition}} statement to throw an \textsf{\small{ArrayIndexOutOfBoundsException}}, which is considered as a type of best practice in the software development process.
Table~\ref{tab:statistics_of_empirical_study} illustrates the experimental results of fine-grained prompting, from which we observe that this strategy solves the three challenges perfectly, with incorrect exception handling and abuse of \textsf{\small{try-catch}} to be completely eliminated. 
Similar to the result of coarse-grained prompting, there still exist three (instead of four) examples that have incomplete exception handling issues, which share the same reason stated previously. 
%because our work shares the same API knowledge base.
One of the coding tasks is solved by ChatGPT accidentally, and the primary cause is the uncertainty of ChatGPT's outputs.
Even the same prompt could trigger different activities of this model, and we will discuss it in Section~\ref{sec:discussion}.

\vspace{-1.5mm}
\begin{tcolorbox}[breakable,width=\linewidth-2pt,boxrule=0pt,top=2pt, bottom=2pt, left=2pt,right=2pt, colback=gray!20,colframe=gray!20]
\textbf{Answer to RQ2:} Fine-grained prompts with specific exceptions and corresponding conditions enable LLMs to identify and handle exceptions accurately, which follows the best practice of the software development process.
\end{tcolorbox}

\begin{table}[t]
\centering
\caption[Table title]{Quality Statistics of Generated Code with Various Prompts}\label{tab:statistics_of_empirical_study}
\vspace{-2mm}
\begin{tabular}
{|c|c|c|cc|}
\hline
\multirow{2}{*}{\textbf{\tabincell{c}{Quality \\ of code}}} & \multirow{2}{*}{\textbf{\tabincell{c}{Direct \\ generating}}} & \multirow{2}{*}{\textbf{\tabincell{c}{General \\ prompting}}} & \multicolumn{2}{c|}{\textbf{\tabincell{c}{Knowledge-driven \\ prompting}}}                               \\ \cline{4-5} 
                                              &                                             &                                             & \multicolumn{1}{c|}{\textit{\textbf{\tabincell{c}{Coarse- \\ grained}}}} & \textit{\textbf{\tabincell{c}{Fine- \\ grained} }} \\ \hline
\textbf{\tabincell{c}{Incomplete \\ exception \\ handling}  }        &                  81                         &           35                                  & \multicolumn{1}{c|}{4}                                 &     3                           \\ \hline
\textbf{\tabincell{c}{Incorrect \\ exception \\ handling} }         &                  4                          &           23                                  & \multicolumn{1}{c|}{0}                                &     0                            \\ \hline
\textbf{\tabincell{c}{Abuse of \\ try-catch}}                   &                  5                          &           20                                  & \multicolumn{1}{c|}{41}                               &      0                          \\ \hline
\textbf{\tabincell{c}{Good \\ practice}  }                        &                  2                          &           14                                  & \multicolumn{1}{c|}{47}                              &     89                           \\ \hline
\end{tabular}
\vspace{1mm}
\scriptsize \emph{\textcolor{red}{*}Note: There is no overlap among the three challenges, and we label them according to a priority of \textbf{incomplete exception handling}, \textbf{incorrect exception handling} and \textbf{abuse of try-catch}.}
\end{table}

\section{The Approach}\label{sec:approach}
\begin{figure*}[t]
  \centering
  \includegraphics[width=1\textwidth]{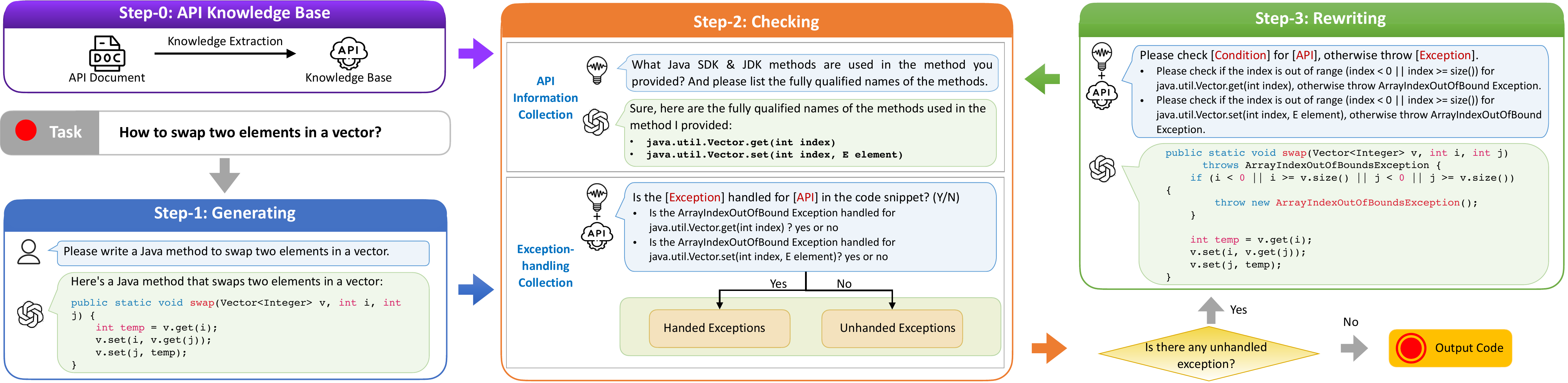}
  \caption{The Overall Framework of \tool-based Code Generation}
  \label{fig:frame}
  \vspace{-3mm}
\end{figure*}
 
\begin{comment}
The overall framework of our approach is illustrated in Figure\ref{fig:frame}.
It consists of two main phrases, i.e., API knowledge base construction and \tool-based code generation. 
Specifically, we first construct an API knowledge base from Java official documentation.
Then, we design a \underline{\textbf{K}}nowledge-driven \underline{\textbf{P}}rompt \underline{\textbf{C}}haining based code generation approach (i.e., \tool) to enhance the code quality with regard to exception handling, which is implemented on ChatGPT. 
\end{comment}

Figure~\ref{fig:frame} illustrates the overall framework of our approach, which consists of one off-line phase and one online phase:
API knowledge base construction (i.e., step-0 in Figure~\ref{fig:frame}) is off-line and \tool-based code generation ( i.e., step-1, step-2 and step-3 in Figure~\ref{fig:frame}) is online.
Specifically, we first construct an API knowledge base from Java official documentation.
Then, we work on top of ChatGPT and design a \underline{\textbf{K}}nowledge-driven \underline{\textbf{P}}rompt \underline{\textbf{C}}haining (\tool)-based code generation approach to improve the code quality with regard to exception handling.

\subsection{API Knowledge Base Construction} 
As investigated in Section~\ref{sec:motivate_prompts}, fine-grained prompts that provide specific exceptions and corresponding conditions can significantly boost the ability of LLMs to generate effective solutions for exception handling, avoiding the fault of incomplete or incorrect handling.
Previous works~\cite{li2018improving, ren2020demystify} demonstrate the feasibility of extracting high-quality knowledge from the official documentation.
We follow such methodologies to mine API exception handling knowledge from Java documentation.
%To obtain fine-grained prompts, we mined API exception-handling knowledge from official documentation, which has been proven to be of high quality in previous studies~\cite{li2018improving, ren2020demystify}.

\subsubsection{Knowledge schema}
We construct the API knowledge base as triples of <entity, relation, entity> and only considered method-level APIs.
The entities have three types of components: \textbf{API}, \textbf{exception} and \textbf{condition}, and the relations between entities include: \textbf{throw} and \textbf{trigger}. 
Specifically, the API knowledge base only has two forms: an API throws exceptions, and a condition triggers an exception.
For example, the task in Figure~\ref{fig:motivation} is related to two APIs in our API knowledge base: \textsf{\small{java.util.Vector.get(int index)}} and \textsf{\small{java.util.Vector.set(int index, E element)}}.
Both of them throw \textsf{\small{ArrayIndexOutOfBoundsException}}, which is triggered by the condition of \emph{``if the index is out of range (index $< 0 $ $||$ index $>=$ size())}''~\cite{specification4get, specification4set}.

\subsubsection{Knowledge extraction}
In this work, we use Java SDK \& JDK API specification~\cite{javadoc} to construct the API knowledge base.
The API documentation is collected from online resources by a web crawling tool~\cite{Beautiful-Soup-4}, and each crawled web page is treated as an API document.
We only keep the semi-structured API declarations and exception-handling specifications, but ignore other contents on the crawled web pages (e.g., code snippets and other textual descriptions), as this work focuses on exception-handling issues brought by API calls.
API declarations describe the API's fully qualified name, which acts as an index of a specific API.
Exception-handling specifications are usually in the form of \emph{``Throw: [Exception] - [Condition].''} and we can use a rule-based method to extract exceptions and conditions from them.
Different from previous works~\cite{li2018improving, ren2020demystify}, which conduct a series of pre-processing such as splitting and tokenizing to the natural language descriptions of the conditions, no further operation is required in this work because of the semantic interpretation capability of LLMs applied by our method.

\begin{comment}
Since we focus on exception-handling issues brought by API calls, we keep the semi-structured API declarations and exception-handling specifications, and ignore other document contents from the crawled web pages (e.g., code snippets, program execution output, and other textual descriptions).  
Due to the high quality of the official documentation, we can easily extract exceptions and conditions from original exception-handling directions, which are usually in the format of \emph{``Throw: [Exception] - [Condition].''}.
For example, the API knowledge base unit in part $C$ of Figure~\ref{fig:motivation} is extracted from \emph{``Throw: ArrayIndexOutOfBoundsException - if the index is out of range (index $< 0 $ $||$ index $>=$ size()).''}
Due to the advanced natural language understanding capabilities of LLMs, traditional preprocessing operations, e.g., splitting and tokenizing, are unnecessary when processing the conditions that trigger exceptions.
\end{comment}

\subsection{\tool-based Code Generation}\label{sec:kpc_approach}
\begin{comment}
In this section, we employ the state-of-the-art (SOTA) code generation technique using a code generator based on ChatGPT, a highly advanced language model, to generate programs from coding tasks described in natural language. 
Subsequently, we use two additional modules, i.e., checking and rewriting, to refine and optimize the generated code.
\end{comment}

Given a natural language described coding task as input, our method takes three steps (i.e., generating, checking-rewriting), which are all supported by prompting the LLM in an AI chain workflow, to output the code with all exceptions being handled (see Figure~\ref{fig:frame}).
The three steps share the same state-of-the-art (SOTA) LLM, namely ChatGPT, and we elaborate the AI chain workflow as shown in Figure~\ref{fig:overall}.

\subsubsection{Generating}
\label{sec:generating}
Typically, a coding task is a sentence of natural language description in the form of ``How to ...''.
The sentence is rephrased into ``Please write a Java method to... '' as a prompt in order to clarify the requirement for LLMs explicitly. 

As one of the most representative techniques to promote the revolution of artificial general intelligence (AGI), ChatGPT performs well on a large variety of natural language as well as software tasks.
For example, previous work~\cite{dong2023self} has demonstrated the ability of ChatGPT to generate high-quality code, inspiring us to take advantage of it to implement the generating step.
This step is responsible for generating the initial version of code according to a given coding task.
Furthermore, the context-aware semantic interpretation of ChatGPT offers the potential for inter-model collaboration and interaction, which will be applied in the checking-rewriting loop.

\begin{comment}
We use GPT-3.5-turbo model~\cite{gpt-3.5-turbo} to implement the generating module, which has been illustrated as the SOTA model for code generation~\cite{dong2023self}.
One of the key advantages of the GPT-3.5-turbo model is its ability to learn from vast amounts of data, allowing it to generate high-quality code snippets with remarkable accuracy. 
% By leveraging this model, we are able to significantly reduce the time and effort required to develop complex software applications.
\xx{Furthermore, the capability of GPT-3.5-turbo model to generate context-aware code also allows more effective interaction among our \tool-based process, enabling them to quickly iterate and refine the code with the assistance of prompts.}
\end{comment}

\subsubsection{Checking}

After the first step of generating, we obtain a piece of code to solve the task (e.g., code of step-1 in Figure~\ref{fig:frame}).
However, there is a high possibility that the code contains defects such as unhandled exceptions, as the input coding task does not provide any domain-specific information about exception handling, which usually appears in API documentation and has been stored in our knowledge base.
We adopt the constructed API knowledge base and formulate a checking step with two parts to collect the exception-handling results.

\begin{itemize}[leftmargin=*, topsep=0pt]
    \item \textbf{API Information Collection.}
    It has been shown that LLMs can achieve good performance on program understanding~\cite{xia2022practical}, which simplifies the process of API extraction.
    To obtain the APIs' fully qualified names, we present the generated code to ChatGPT and ask it with the prompt ``What Java SDK \& JDK  methods are used in the method you provided? Please list the fully qualified names of the methods.''
    This API information collection part of step-2 in Figure~\ref{fig:frame} illustrates an example of how to extract the two API fully qualified names, i.e., \textsf{\small{java.util.Vector.get(int index)}} and \textsf{\small{java.util.Vector.set(int index, E element)}}, from the generated code.
    We also use the extracted API names to construct links with the API knowledge base for further analysis.
    
    \item \textbf{Exception-handling Collection.}
    We first create a prompt template ``Is the \textcolor{red}{[Exception]} handled for \textcolor{red}{[API]} in the code snippets? (Y/N)'', in which the [Exception] and [API] are placeholders.
    Then the prompt template is instantiated by filling placeholders with related APIs and their corresponding exceptions according to information in the API knowledge base.
    For example, the prompt template in Figure~\ref{fig:frame} (step-2) is instantiated into two sub-questions, i.e., ``Is the ArrayIndexOutOfBoundException handled for java.util.Vector.get(int index)?'' and ``Is the ArrayIndexOutOfBoundException handled for java.util.Vector.set(int index, E element)?''.
    Similar to the previous part (i.e., API information collection), we provide ChatGPT with code and ask it questions with the aim of acquiring the exception-handling information.
    This operation is conducted on each extracted API, and all the handled/unhandled exceptions are collected in this step.
\end{itemize}

\subsubsection{Rewriting}
\begin{comment}
This module asks LLMs to rewrite the generated code with the prompt template ``Please check \textcolor{red}{[Condition]} for \textcolor{red}{[Method]}, otherwise throw \textcolor{red}{[Exception]}.'', if there are unhandled exceptions collected in the checking module.
The prompt template should also be instantiated with API knowledge base by replacing \textcolor{red}{[Condition]}, \textcolor{red}{[Method]} and \textcolor{red}{[Exception]}.
For example, using our \tool-based approach to rewrite the generated code in Figure~\ref{fig:motivation}, we instantiate the prompt template into ``Please check if the index is out of range (index < 0 || index >= size()) for java.util.Vector.get(int index), otherwise throw ArrayIndexOutOfBoundsException'' and ``Please check if the index is out of range (index < 0 || index >= size()) for java.util.Vector.set(int index, E element), otherwise throw ArrayIndexOutOfBoundsException''. 

\xx{[Where to put??]}
In \tool-based code generation, the checking and rewriting modules are iterative. 
Specifically, the code rewritten should be checked to collect exception-handling information and will be rewritten continually until all related exceptions in the knowledge base are handled.
\end{comment}
This step is technically similar to generating and checking, which applies the proper prompt to make ChatGPT generate the required code.
The main difference between them is the prompt creation strategy.
Specifically, we design a new prompt template ``Please check \textcolor{red}{[Condition]} for \textcolor{red}{[API]}, otherwise throw \textcolor{red}{[Exception]}.'' for this step.
If there exist any unhandled exceptions in the code collected in step-2, the prompt template is instantiated with the information from the API knowledge base.
We integrate all the exceptions into one prompt and ask ChatGPT to handle the exceptions at one time and rewrite the code.
For example, in Figure~\ref{fig:frame} (Rewriting), we instantiate the prompt template into ``Please check if the index is out of range (index < 0 || index >= size()) for java.util.Vector.get(int index), otherwise throw ArrayIndexOutOfBoundsException. Please check if the index is out of range (index < 0 || index >= size()) for java.util.Vector.set(int index, E element), otherwise throw ArrayIndexOutOfBoundsException''.

The checking-rewriting steps will keep going iteratively until no more unhandled exception remains in the generated code.

\section{Experimental Setup}
\subsection{Research Question}
 In the evaluation, we study the following research questions:
\begin{itemize}[leftmargin=*, topsep=0pt]
\item \textbf{RQ3: How effective is our \tool in identifying and handling exceptions during code generation tasks?}
To answer this question, we perform a detailed statistical analysis to determine the number of potential exceptions that may arise in the generated code, as well as the number of checking-rewriting loops required by \tool to address these exceptions.

\item \textbf{RQ4: To what extent can our \tool effectively assist in handling exceptions in LLM code generation?} We compare our approach with the state-of-the-art (SOTA) code generator, i.e.,ChatGPT, and employ both static and dynamic evaluation methods to assess the improvements in exception-handling issues in the generated code.
\end{itemize}

\subsection{Task Collection and Implementation}
The coding tasks of our experiments are collected from Java SDK \& JDK API specification~\cite{javadoc}.
Following previous work~\cite{huang2021characterizing, mastropaolo2023robustness}, we also use the API functional descriptions as the coding tasks. 
However, not all tasks are considered in this work.
We only considered the tasks that LLMs can generate code using Java SDK \& JDK APIs with potential exceptions. 
Thus, by analyzing code generated by LLMs, we collect 3,079 Java coding tasks from 11,259 API functional descriptions.

Both our \tool-based code generation approach and our baseline (i.e., ChatGPT) are implemented on top of the GPT-3.5-turbo model~\cite{gpt-3.5-turbo}, which has been proven to achieve the best performance in code generation~\cite{dong2023self}.

%gpt-3.5-turbo
%experiment environment

\section{Evaluation}\label{sec:evaluation}

\subsection{Efficiency of \tool (RQ3)}
\subsubsection{Method}
Our \tool-based code generation approach is an iterative process, it is crucial to evaluate its efficiency in dealing with exception-handling issues. 
As our approach is implemented on top of model-3.5-turbo, whose response time cannot be artificially controlled, the best way to assess our approach's efficiency is to measure the number of exceptions that must be addressed in each loop during exception handling, and how many loops are necessary to complete the exception handling tasks.

Note that the initial round of checking-rewriting steps is considered as the first loop. Additionally, it's important to highlight that various exceptions that arise from the same API are identified as distinct exceptions.

\subsubsection{Result}
% \begin{figure}[t]
%   \centering
%   \includegraphics[width=0.5\textwidth]{pic/fig3.png}
%   \caption{The Distribution of Checking-Rewriting Loops}
%   \label{fig:loop}
% \end{figure}

% \begin{figure}[t]
%   \centering
%   \includegraphics[width=0.5\textwidth]{pic/fig4.png}
%   \caption{The Distribution of Exceptions}
%   \label{fig:exception}
% \end{figure}

\begin{figure}[ht]
  \centering
  \vspace{-3mm}
  \begin{minipage}[t]{0.48\linewidth}
    \centering
    \subfigure[\tiny Distribution of Checking-Rewriting Loops]{ \label{fig:loop}
    \includegraphics[width=1\textwidth]{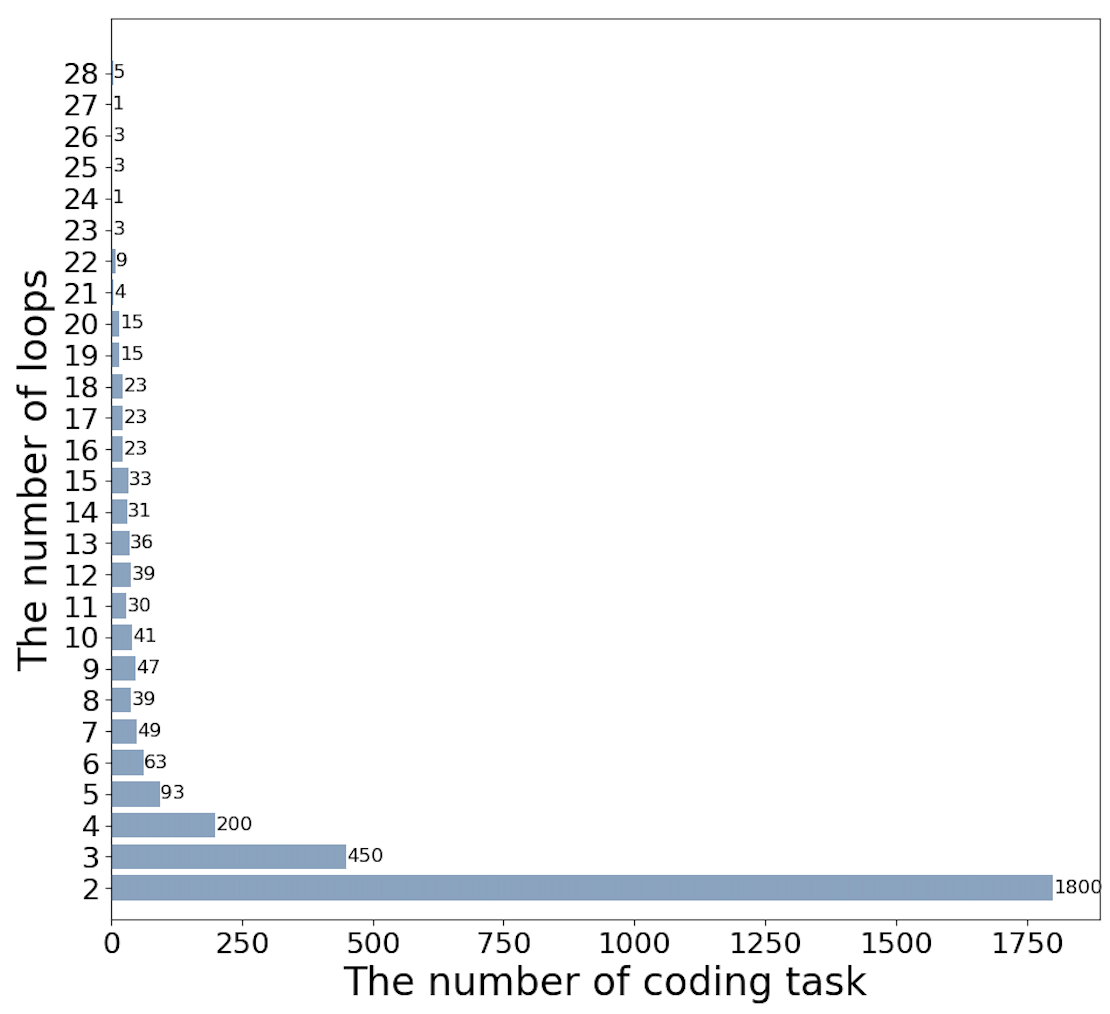}}
  \end{minipage}
  \hspace{0.01in}
  \begin{minipage}[t]{0.48\linewidth}
    \centering
   \subfigure[\tiny Distribution of Unhandled Exceptions]{\label{fig:exception}
    \includegraphics[width=1\textwidth]{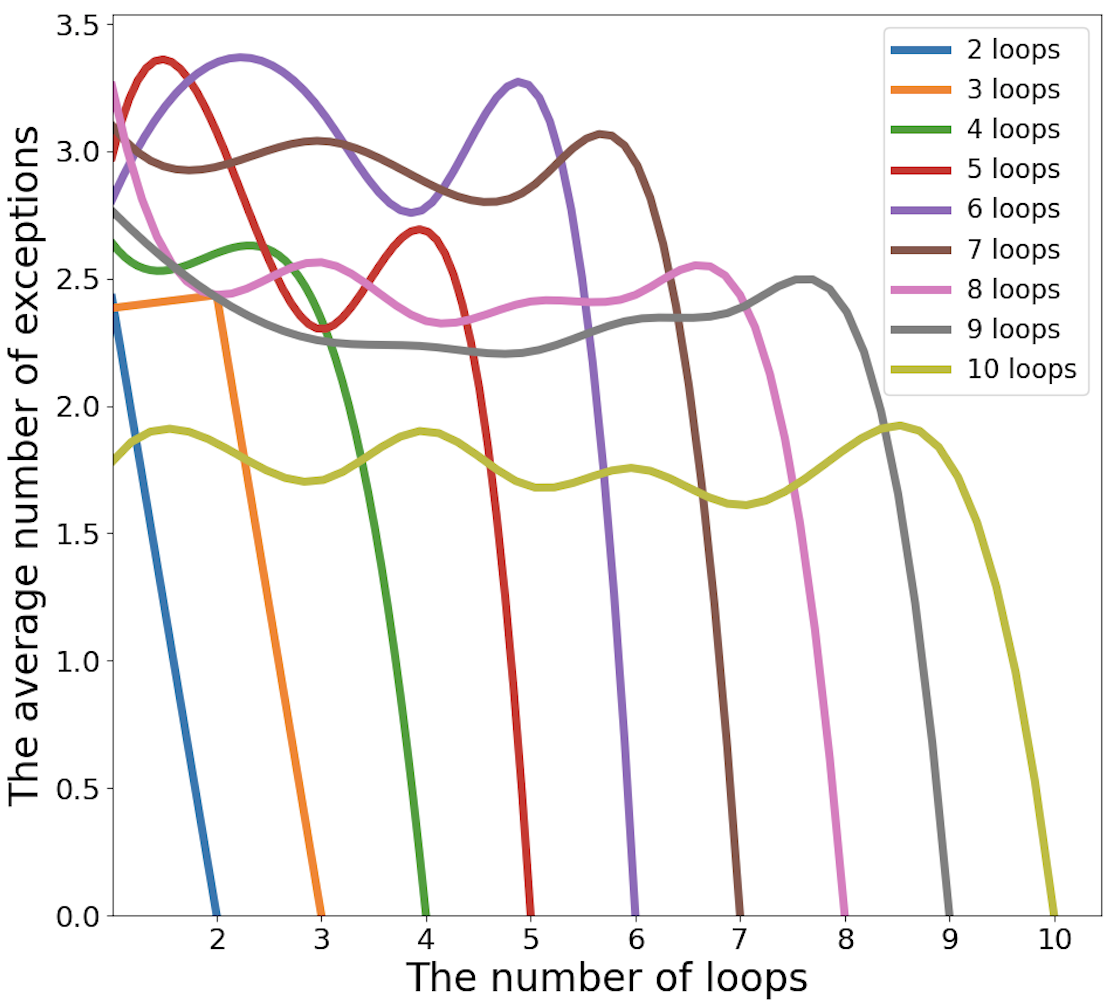}}
  \end{minipage}
  \vspace{-3mm}
  \caption{Distributions of \tool-based Code Generation}
  \label{fig:subfigure-demo2}
\end{figure}

Figure~\ref{fig:loop} shows the distribution of checking-rewriting loops for all coding tasks completed by our \tool-based code generation approach.
In our experiment, the number of checking-rewriting loops required ranges from a minimum of 2 to a maximum of 28. 
Among the 3,079 coding tasks, 90.35\% (2,782) tasks could be completed within 10 loops, and 58.46\% (1,800) tasks require only two loops for completion.
In this case, the results suggest that exceptions in the majority of coding tasks can be efficiently handled by our \tool within a limited number of checking-rewriting loops.

%average exception computation
%exception：1-3
%5，6 ->0
Figure~\ref{fig:exception} displays the average number of unhandled exceptions collected within each loop, which refers to the average number of unhandled exceptions per loop across all tasks that can be completed in the same number of loops.
It is obvious that the average number of unhandled exceptions per loop ranges from $1.61$ to $3.35$, which indicates that each loop is simple and not required to handle too many exceptions.

We also observe that the average number of unhandled exceptions per loop can fluctuate (e.g., 5 loops and 6 loops in Figure~\ref{fig:exception}), which is because once the generated code is rewritten, new unhandled exceptions may be introduced. 
However, all unhandled exceptions can be resolved by the end of the loop, resulting in zero unhandled exceptions upon completion.

\vspace{-1.5mm}
\begin{tcolorbox}[breakable,width=\linewidth-2pt,boxrule=0pt,top=2pt, bottom=2pt, left=2pt,right=2pt, colback=gray!20,colframe=gray!20]
\textbf{Answer to RQ3:} Our \tool-based code generation approach is designed to efficiently handle the majority of exceptions using a small number of checking-rewriting loops, typically within 10 simple loops.
\end{tcolorbox}

\subsection{Effectiveness of \tool (RQ4)} \label{sec:rq4}
\subsubsection{Method}
To evaluate the effectiveness of ChatGPT and our \tool-based approach in exception handling from both static and dynamic perspectives, we select three evaluation methods: LLMEva, CodeReview, and EvoSuite.
LLMEva and CodeReview use static code analysis to evaluate the generated code, and EvoSuite generates test cases to make the dynamic validation.

% We select three evaluation methods (i.e., LLMEva, CodeReview and EvoSuite) to measure the performance of exception handling in the code generated by ChatGPT and our \tool-based approach from both static and dynamic aspects.
% Specifically, LLMEva and CodeReview belong to static methods using static code analysis, and EvoSuite is a dynamic method that generates test cases to find runtime bugs in the generated code.

Details of the three evaluation methods are as follows:
\begin{itemize}[leftmargin=*, topsep=0pt]
\item \textbf{LLMEva:} 
It leverages LLMs' capability of comprehending code to evaluate the performance of generated code.
Specifically, LLMEva evaluates the exception-handling performance of the generated code by posing the prompt to ChatGPT: ``Can the code handle all exceptions in good practice? (Y/N)?''. 

\item \textbf{CodeReview}: 
It is a manual code review method.
Two of the authors independently label whether all exceptions in the generated code are handled in accordance with good practice. 
After completing respective labels, they compared their assessments and discussed any disagreements to arrive at a consensus. 
Note this process was conducted by the same authors who have conducted the empirical study before, which adds to the reliability and validity of our results.

\item \textbf{EvoSuite~\cite{fraser2011evosuite}:} 
It is a state-of-the-art search-based software testing approach~\cite{harman2012search,fraser2011evosuite} used for generating test cases for Java code, and it has been demonstrated to achieve high code coverage and improve bug detection capabilities.
Generating test cases for code is an effective way to assess the quality of the code and ensure that it can avoid errors through appropriate exception handling. 
Many search algorithms have been proposed for EvoSuite, and we choose DynaMOSA~\cite{panichella2017automated} as the search algorithm, which optimizes multiple coverage targets simultaneously.
Search algorithms have various parameters to set, but previous work~\cite{arcuri2013parameter} shows that parameter tuning SBST is extremely expensive and not necessary compared to default parameter values.
Thus, we use the default settings of DynaMOSA.
Moreover, considering our work focuses on exception handling in the generated code, we choose branch coverage and exception coverage as the optimization objective.
To ensure the efficient allocation of resources, we established a time budget of two minutes per class for this process and repeated it 10 times. 
The executions is run on gnu/Linux system (Ubuntu
18.04.6 LTS) with 5.4.0-128-generic Linux kernel, 28-core 2.60GHz
Intel(R) Xeon(R) Gold 6348 CPU and 1TB RAM.  
\end{itemize}

For LLMEva, we evaluate results of all coding tasks in our experiment because the evaluation process is automated. 
While to tackle the time-consuming nature of both CodeReview and EvoSuite, we employ a statistical sampling method~\cite{singh2013elements} to analyze $MIN$ randomly selected instances of the coding tasks. 
$MIN = 384$ in this work, which ensures the estimated accuracy is in 0.05 error margin at 95\% confidence level.

\subsubsection{Result}

\begin{table}
\centering
\caption{The number of code with good exception-handling practices}
\label{tab:overall_performance}\vspace{-2mm}
\resizebox{0.48\textwidth}{!}{
\begin{tabular}{|c|c|c|c|c|}
\hline
                    & \textbf{\# of Tasks} & \textbf{ChatGPT} & \textbf{\tool}  & \textbf{Improve} \\ \hline
\textbf{LLMEva}    & 3,079                & 1,056 (\textbf{34.30\%})    & 2,195(\textbf{71.29\%}) & 107.86\%       \\ \hline
\textbf{CodeReview} & 384                 &     56 (14.58\%)               &        380 (98.86\%) &  
 578.57\%\\ \hline
\textbf{EvoSuite}   & 384                 &    353 (\textbf{91.93\%})            & 371 (\textbf{96.61\%})   & $\downarrow$ 18 bugs  \\ \hline          
\end{tabular}
}
\end{table}

Table~\ref{tab:overall_performance} shows the number of code with good exception-handling practices generated by our \tool and ChatGPT, which can reflect their effectiveness of exception handling under different evaluation methods.
Overall, our \tool demonstrates significant advantages across all three evaluation methods, with exception handling improvements of 107.86\%, 578.57\%, as well as a reduction of 18 real bugs, respectively.

According to LLMEva, 34.30\% (1,056 out of 3,079) of the code generated by ChatGPT and 71.29\% (2,195 out of 3,079) of the code generated by our \tool-based approach are proven to be with good exception-handling practice.
As discussed in Section~\ref{sec:motivate_challenges}, LLMs have several limitations in handling exceptions. 
The reliability of results evaluated by LLMEva is uncertain. 
% However, a comparison of LLMEva results with code generated directly by ChatGPT (71.29\% $V.S.$ 34.30\%) indicates that our \tool-based approach yields significant improvements in exception handling.
% Due to the many shortcomings of LLMs in handling exceptions (see Section~\ref{sec:motivate_challenges}), the reliability of the results evaluated by LLMEva remains uncertain. 
However, based on the relative results (71.29\% vs. 34.30\%), it is evident that our \tool-based approach exhibits a significant improvement in exception handling compared to the code generated directly by ChatGPT.

The same results are more pronounced in manual code review.
According to code review (i.e., the ``CodeReview'' row in Table~\ref{tab:overall_performance}), only 14.58\% of code generated by ChatGPT can meet good exception handling practice, while that of our \tool-based approach is as high as 98.86\%. 
For manual code review, the Cohen's Kappa between the two authors is 0.98, which indicates they are with a significantly high agreement in the labeling results.

With careful observation, we find the main reason for the huge gap between the results of CodeReview and LLMEva is LLMs lack sufficient exception-handling knowledge and best practices for exception handling.
Specifically, LLMs prefer to use general exceptions handling strategies, e.g, ``\textsf{\small{RuntimeException}}'', instead of specific exceptions, e.g., ``\textsf{\small{ArrayIndexOutOfBoundsException}}''.
Using general exception handling can catch and handle multiple types of exceptions with a single catch block, but may result in the loss of important information needed to diagnose and fix specific errors, which is considered a bad practice in this work. 
% In addition to using specific exceptions, our \tool-based approach uses fine-grained knowledge-driven prompts consisting of specific exceptions and corresponding conditions, which can help LLMs generate more informative and accurate error messages, which can be a valuable tool for developers when debugging code.
By contrast, our \tool-based approach leverages fine-grained knowledge-driven prompts that include specific exceptions and their corresponding conditions.
Our approach enhances the ability of LLMs to generate informative and accurate error messages. 
By providing developers with more detailed error messages, our approach can be a valuable tool for debugging.

Furthermore, the results of EvoSuite 
% According to static analysis, our \tool-based approach can perform much better than ChatGPT in exception handling, and then we used Evosuite to dynamically validate whether the code generated by ChatGPT and our \tool can correctly handle exceptions in the runtime.
% Table~\ref{tab:overall_performance} 
reveals that out of the 384 randomly selected tasks, a total of 31 ($384-353$) unique bugs across 10 runs can be detected in the code generated by ChatGPT. 
In comparison, 13 ($384-371$) unique bugs are detected in the code generated by our \tool-based approach.
As the ultimate goal of exception handling is to minimize runtime errors, results from EvoSuite indicate that our \tool-based approach has the potential to help developers handle real bugs in their code.
Although our approach only reduces 18 more real bugs in the sampled dataset compared with EvoSuite, we would like to remind the readers that EvoSuite is a sophisticated software tool that requires long-term research and significant amount of enginering effort to build and maintain.
In contrast, by standing on the shoulder of the LLMs and through the design of an AI chain and knowledge-driven prompts, which is much simpler than EvoSuite, our approach achieves the performance on par with EvoSuite.
This result sheds the light on the new opportunities to build software analysis tools on top of the LLMs.

\vspace{-1.5mm}
\begin{tcolorbox}[breakable,width=\linewidth-2pt,boxrule=0pt,top=2pt, bottom=2pt, left=2pt,right=2pt, colback=gray!20,colframe=gray!20]
\textbf{Answer to RQ4:} Our \tool-based approach can enhance the ability of LLMs to generate informative and accurate error messages, which can also help reduce runtime bugs in the generated code.
\end{tcolorbox}

\section{User Study}\label{sec:userstudy}
We conduct a user study to evaluate the usefulness of our \tool-based code generation approach for helping developers write code efficiently and correctly.

% \usepackage{multirow}
\begin{table*}[]
\caption{Coding Tasks and Results of User Study}\label{tab:userstudy} \vspace{-2mm}
\resizebox{1\textwidth}{!}{
\begin{tabular}{|c|p{0.3\linewidth}|c|ccc|ccc|cc|}
\hline
\multirow{2}{*}{\textbf{No.}} & \multicolumn{1}{c|}{\multirow{2}{*}{\textbf{Coding Task}}}                                                                       & \multirow{2}{*}{\textbf{Difficulty}} & \multicolumn{3}{c|}{\textbf{Correctness}}                                            & \multicolumn{3}{c|}{\textbf{\begin{tabular}[c]{@{}c@{}}Time Consumption \\ (second)\end{tabular}}} & \multicolumn{2}{c|}{\textbf{Usefulness}}         \\ \cline{4-11} 
                              & \multicolumn{1}{c|}{}                                                                                                            &                                      & \multicolumn{1}{c|}{\textit{G-1}} & \multicolumn{1}{c|}{\textit{G-2}} & \textit{G-3} & \multicolumn{1}{c|}{\textit{G-1}}      & \multicolumn{1}{c|}{\textit{G-2}}      & \textit{G-3}     & \multicolumn{1}{c|}{\textit{G-2}} & \textit{G-3} \\ \hline
T1                            & Please write a Java method that removes the char at the specified position in this sequence.                                     & Medium                               & \multicolumn{1}{c|}{1/4}          & \multicolumn{1}{c|}{3/4}          & 4/4          & \multicolumn{1}{c|}{84.5}              & \multicolumn{1}{c|}{267.5}             & 131.3            & \multicolumn{1}{c|}{3.5}          & 4.0          \\ \hline
T2                            & Please write a Java method to swap two elements in a vector using Java                                                           & Easy                                 & \multicolumn{1}{c|}{2/4}          & \multicolumn{1}{c|}{2/4}          & 4/4          & \multicolumn{1}{c|}{82.3}              & \multicolumn{1}{c|}{225.0}            & 148.8            & \multicolumn{1}{c|}{3.0}          & 4.0          \\ \hline
T3                            & Please write a Java method that acquires a lock on the given region of this channel's file.                                      & Hard                                 & \multicolumn{1}{c|}{1/4}          & \multicolumn{1}{c|}{1/4}          & 2/4          & \multicolumn{1}{c|}{500.0}             & \multicolumn{1}{c|}{385.5}             & 420.0            & \multicolumn{1}{c|}{3.5}          & 3.8          \\ \hline
T4                            & Please write a Java method that creates a new instance of URLClassLoader for the specified URLs and default parent class loader. & Hard                                 & \multicolumn{1}{c|}{0/4}          & \multicolumn{1}{c|}{1/4}          & 2/4          & \multicolumn{1}{c|}{253.0}             & \multicolumn{1}{c|}{248.0}               & 480.0            & \multicolumn{1}{c|}{3.8}          & 3.5          \\ \hline
T5                            & Please write a Java method that interrupts a running Thread in Java.                                                              & Medium                               & \multicolumn{1}{c|}{1/4}          & \multicolumn{1}{c|}{2/4}          & 4/4          & \multicolumn{1}{c|}{159.0}             & \multicolumn{1}{c|}{147.3}             & 190.0            & \multicolumn{1}{c|}{3.8}          & 3.5          \\ \hline
T6                            & Please write a Java method that split a string into a number of substrings in Java                                               & Easy                                 & \multicolumn{1}{c|}{2/4}          & \multicolumn{1}{c|}{2/4}          & 2/4          & \multicolumn{1}{c|}{95.3}              & \multicolumn{1}{c|}{180.5}             & 107.5            & \multicolumn{1}{c|}{3.3}          & 2.8          \\ \hline
Average                        & \multicolumn{1}{c|}{-}                                                                                                           & -                                    & \multicolumn{1}{c|}{29.17\%}       & \multicolumn{1}{c|}{45.83\%}       & \textbf{75.00\%}          & \multicolumn{1}{c|}{\textbf{195.7}}             & \multicolumn{1}{c|}{242.3}             & 246.3            & \multicolumn{1}{c|}{3.5}          & \textbf{3.6}          \\ \hline
\end{tabular}
}\vspace{-5mm}
\end{table*}

\subsection{User Study Design}

\subsubsection{Tasks and Procedure}
\begin{comment}
Table~\ref{tab:userstudy} shows the coding tasks of the user study, which are selected from the dataset of empirical study (Section~\ref{sec:empirical}) and evaluation with regard to the various difficulties of coding tasks.
Based on our own analysis of these tasks, we estimate that T2/T6 are easy, T1/T5 are medium, and T3/T4 are difficult.

Our study involves asking all participants to write Java code for given tasks. 
They are allowed to use online resources such as API documentation and Stack Overflow, but they cannot use any code generators to complete the tasks.
Each task has a time limit of up to ten minutes, and the entire study is expected to be completed within an hour. 
We also conduct interviews with each participant to gain a better understanding of their thought processes behind their responses to specific tasks. 
This will help us collect valuable insights on how programmers approach problem-solving tasks and make decisions when writing code.
\end{comment}

As shown in Table~\ref{tab:userstudy}, we select six typical Java programming tasks from the dataset of empirical study (Section~\ref{sec:empirical}) and evaluation (Section~\ref{sec:evaluation}), which includes two easy tasks (T2 and T6), two medium tasks (T1 and T5) and two hard tasks (T3 and T4).
The participants are expected to use a certain kind of Java API to write code and implement a method that satisfies the requirement of each task. 
It is free to access any online resources, including API documentation and Q\&A forums, except the code generator (e.g., CodeX and ChatGPT).
Each task has a time limit of ten minutes, which means the user study can be finished within one hour.
After the user study, we conduct a short survey by collecting feedback on each participant, which is used to understand their behaviors and improve our work in the future.

% In our study, we ask all participants to write down Java code according to the given task.
% The participants can access online materials (e.g., API documentation, Stack Overflow etc..) but cannot use any code generators to help complete the tasks.
% Each question is allocated up to ten minutes, and the whole study completes in about an hour. 
% We also interview each participant to collect the rationale behind his/her answers to certain tasks.

\subsubsection{Participants}
We recruit 12 participants from an IT company that has over 2,000 developers to attend the user study.
The participants have 1 to 5 (on average of 3.2) years of Java development experience on both commercial and open-source projects.
They are divided into three groups (i.e., G-1, G-2, G-3) equally, and each group consists of four participants.
\begin{comment}
We recruit 12 developers from an IT company that has over 2,000 developers. 
These 12 developers have 1 to 5 years (on average 3.2 years) of Java development experience on either commercial or open-source projects. 
Based on the development experience of these developers, we divide them into three ``comparable'' groups: G-1, G-2 and G-3. 
Each group has 4 developers. 
\end{comment}
G-1 and G-2 are control groups, where G-1 is given only the coding tasks, while G-2 is given both the coding tasks and reference code generated by ChatGPT.
G-3 is the experimental group that is given both the coding tasks and reference code generated by our \tool-based code generation approach.

\subsubsection{Evaluation Metrics}
%Our evaluation of participant performance is based on two key factors: task completion time and answer correctness. 
We consider two key factors for evaluating the performance of participants: task completion time and answer correctness. 
Task completion time reflects how fast a participant can complete a coding task.
Answer correctness represents whether the code submitted by a participant is actually an appropriate solution with good exception-handling practice to the task.
%The two authors collaboratively determine the correctness of each submitted code by examining whether it can solve the task with good exception-handling practice. 
The two authors first review the code to determine the correctness of it independently and discuss for final decisions if there exist different results.
We compute Fleiss's Kappa~\cite{fleiss1971measuring} to examine the agreement between the two authors.
If the submitted code is annotated as correct, the participant gets 1 mark, otherwise 0 mark. 
We use Wilcoxon signed-rank test~\cite{wilcoxon1992individual} with Bonferroni correction~\cite{weisstein2004bonferroni} to determine if the performance variation across different groups is statistically significant.
For example, if the corresponding Wilcoxon signed-rank test result (i.e., p-value) is less than 0.05, we can consider one group performs better than the other at the confidence level of 95\%. 

\begin{comment}
Furthermore, for participants in G-2 and G-3, we also ask two supplementary questions for each task regarding the reference code they were given. 
Specifically, we ask Q1:How effectively does the reference code handle exceptions? Does it handle them incompletely, incorrectly, or misuse the try-catch mechanism? Q2: To what extent did the reference code aid you in completing the task successfully?
For Q1, if the participant can correctly point out the exception handling issues in the reference code, he/she will get 1 mark, otherwise 0 mark. 
These supplementary questions help us gain a better understanding of how the participants interacted with the reference code, and whether it was helpful or hindering to their problem-solving process. 
This information will provide valuable insights into the effectiveness of reference code in assisting programmers in solving complex programming tasks.
\end{comment}

Furthermore, we ask participants in G-2 and G-3 a supplementary question, ``To what extent does the reference code help you to complete the task successfully?''.
They are supposed to answer it by giving a score of 0 to 4 for each task, where 0 means helpless, and 4 means very helpful.

\subsection{Result}
Table~\ref{tab:userstudy} shows the results of the user study. 
\subsubsection{Correctness}
The ``Correctness'' column in Table~\ref{tab:userstudy} indicates the accuracy of each task completed by different groups. 
For instance, a value of ``1/4'' means that only one out of four participants completed the coding task correctly.

We can see from Table~\ref{tab:userstudy} that without any reference to complete the coding tasks, G-1 only achieves an average correctness of 29.17\%.
%From the ``Average'' row, Group G-1, which completes the coding tasks without any reference code, achieve an average correctness rate of 29.17\%.
All coding tasks have limited numbers of correct answers, with only one or two participants being able to complete them correctly.
In contrast, G-2 and G-3 have reference code to complete the tasks, which obtain relatively higher average correctness of 45.83\% and 75.00\%, respectively.
This suggests the reference code can significantly assist participants in completing coding tasks, especially when the code is generated by \tool-based approach and provide sufficient information about exception handling.
%Especially the reference code generated by our tool-based approach was more effective in reminding participants of the necessary exceptions to handle, when compared to the reference code generated by ChatGPT.
The Wilcoxon signed rank test shows that the differences of the correctness between G-1 and G-2, G-1 and G-3, and G-2 and G-3 are statistically significant at $p-value < 0.05$.

%When evaluating the correctness of each coding tasks, there are several observations to consider. 
We gain the following research findings when evaluating the correctness of coding tasks.
In cases of no reference code is provided (i.e., G-1), easy tasks such as T2 and T3 are more likely to be completed correctly, while hard tasks such as T3 and T4 are still big challenges for the participants to solve. 
However, the situation becomes more complex when a reference code is provided.
In G-2, the reference code generated by ChatGPT may contain numerous problems with exception handling, which could mislead participants. 
%Consequently, there may be no improvements for many tasks compared to G-1, including T2, T3, and T6.
Consequently, there is no improvement for half of the tasks, including T2, T3 and T6.
On the other hand, the reference code generated by our \tool-based approach can provide more useful information for exception handling, achieving significant improvements in correctness.
Specifically, all four participants successfully completed tasks T1, T2, and T5, showcasing the effectiveness of our approach.
%For instance, all the four participants in G-3 completed tasks T1, T2, and T5 correctly.
It is worth noting that even provided with reference code, participants in G-2 and G-3 may occasionally complete T6 incorrectly. 
We find the main reason is that both ChatGPT and our \tool fail to generate the correct reference code as the API documentation does not include exception-handling specifications for the APIs used in this task, which has been explained in Section~\ref{sec:motivate_prompts}. 

\subsubsection{Time Consumption}
\begin{comment}
The ``Time Consumption'' column of Table~\ref{tab:userstudy} refers to the average completion time of the group for each task.
The G-1 participants take the shortest time to complete the tasks (i.e., with an average of 195.7 seconds), the G-2 and G-3 participants take similar time (i.e. with average values of 242.3 seconds and 246.3 seconds, respectively).

Participants in G-1 tend to complete the coding tasks on their own without relying heavily on the provided reference code.
As a result, they can complete the tasks more quickly than participants in G-2 and G-3.
In contrast, G-2 and G-3 participants tend to be more cautious when working with the reference code. 
They may spend more time checking whether the reference code can be helpful in completing the tasks correctly.
While reference code may take an extra ten seconds or so, it is unlikely to cause significant problems for developers, especially when compared to the importance of accuracy in coding tasks.
The Wilcoxon signed-rank test shows the differences of the task completion time between G-1 and G-2, G-1 and G-3, and G-2 and G-3 are statistically significant at $p-value < 0.05$.
\end{comment}

The ``Time Consumption'' in Table~\ref{tab:userstudy} refers to the meantime that a group completes each task, in which G-1 takes the shortest time to finish the tasks with an average of 195.7 seconds.
We find that the participants in G-1 tend to complete the coding tasks on their own without referring to any generated code.
In contrast, participants in G-2 and G-3 are more likely to read the reference code carefully to extract useful information about the tasks, which costs them more time to finish the tasks, with average values of 242.3 seconds and 246.3 seconds, respectively.
% There is no significant difference in the time consumption between the two groups since both ChatGPT and our \tool generate similar lengths of code.
Although it takes about 25\% more time to read the reference code, it is worth that the participants have more chances to solve the coding task correctly.
The Wilcoxon signed-rank test shows the differences in the task completion time between G-1 and G-2, G-1 and G-3, and G-2 and G-3 are statistically significant at $p-value < 0.05$.

\subsubsection{Usefulness and Interview}
%The ``Usefulness'' column of Table~\ref{tab:userstudy} is a unique feature of G-2 and G-3. 
%It represents the participants' perceptions of using code generators to aid them in programming. 
%This column is scored on a scale of 0 to 4, with 0 indicating that the generated code is useless and 4 indicating that it is very useful.
\begin{comment}
Participants in both G-2 and G-3 generally find the provided reference code to be useful, giving it an average score of over 3.5. 
However, when asked about the exception-handling solutions in the reference code, most participants in both groups were unable to identify problems accurately.
What's worse, from the output code, we also find participants can be easily misled by the reference code, as many participants in G-2 output the error code the same as the given reference code. 
This indicates that they may not be proficient at handling exceptions or may not have sufficient knowledge of exception handling, despite some of them being senior Java developers.

Hence, in cases where participants are not proficient at handling exceptions, our \tool-based approach can automatically prompt LLMs in exception handling, which is extremely beneficial for the efficiency and effectiveness of code generation.
\end{comment}

The ``Usefulness'' in Table~\ref{tab:userstudy} is a unique feature of G-2 and G-3, and we obtain similar scores of 3.5 and 3.6 for the two groups, respectively, which means the participants believe the code generated by both ChatGPT and \tool is helpful.
We further analyze the interview records and have two interesting findings.
Firstly, most of the developers have weak awareness of exception handling.
For example, when we ask them ``Is there any exception handling issues in the reference code?'', they feel very hard to identify the problem accurately, even for those APIs that are frequently used.
This phenomenon applies not only to junior developers but also to senior developers, who often overlook seemingly ``minor'' issues.
That is precisely why developers can inadvertently make mistakes, and even impeccably designed software may still contain bugs.
% Secondly, ChatGPT is exceptionally skilled at misleading developers to write code.
% For example, some participants firmly believe that the ChatGPT generated code is a good solution for the coding task, even though there are exception-handling issues in the code. 
% This is because ChatGPT is trained by large amount of online resources including open-source code, which is the common practice of software development community.
Second, ChatGPT is very ``good'' at generating flawed code that appears to be correct, thereby misleading developers.
That's because ChatGPT was trained with a plethora of online resources, and the answers it generates partly reflect common practices in the software development community, but common practices aren't necessarily the right ones.
Hence, our reliable knowledge-driven approach is therefore of paramount importance here.

\section{Discussion} \label{sec:discussion}
\subsection{Limitations of \tool}
As our \tool-based code generation approach is implemented on top of ChatGPT, several limitations need acknowledging.

\subsubsection{Code understanding}
ChatGPT's capability to understand code is limited because it is primarily designed for natural language processing and lacks direct experience with programming languages.
Generally, ChatGPT is able to recognize and generate basic code syntax, e.g., variable declarations and conditional statements, but does not have a deep understanding of underlying concepts and semantic interpretation of programs.
Therefore, writing comprehensive code to solve complex coding tasks remains a big challenge for ChatGPT.
%While ChatGPT can recognize and generate some basic code syntax, e.g., variable declarations and conditional statements, it may not have a deep understanding of the underlying concepts and principles of programming. 
%Therefore, it may not be able to provide accurate or comprehensive answers to complex coding questions or perform advanced programming tasks.

In this work, leveraging prompts to interact with ChatGPT exists in most steps of our approach, e.g., collecting API and exception-handling information in step-2 (see Section~\ref{sec:kpc_approach}).
ChatGPT's limitation in code understanding is one of the potential negative impacts on the performance of this work.

\subsubsection{Trade-off between efficiency and effectiveness}
Our \tool interacts with ChatGPT iteratively to deal with the exception handling issues in the code until the generated code is with good practice.
According to the CodeReview (Section~\ref{sec:rq4}) and user study (Section~\ref{sec:userstudy}), we also realize that the more iterations \tool interacts with ChatGPT, the more complex the generated code will be, resulting in confusing and difficulties for developers to understand the code.
Although increasing loops give our tool a higher probability of solving all exceptions in the coding tasks, this strategy ignores the requirement of efficiency and might not be a good practice in the industry.
As demonstrated in RQ3, a reduced loop number within 10 loops is suggested for balancing efficiency and effectiveness.

\subsection{Threats to Validity}

\subsubsection{Internal Validity} 
There might be inaccuracy when labeling the exception-handling practice for generated code (i.e., Section~\ref{sec:rq4}) and scoring for users' answers (i.e., Section~\ref{sec:userstudy}).
Both annotators have more than five years of programming experience in Java programming. 
Additionally, two annotators check all the information about a vulnerability independently and discuss if they cannot reach an agreement.

\subsubsection{External Validity}
One of the threats is the exception-handling specifications in Official API documenation~\cite{javadoc} are not comprehensive, which makes our \tool-based approach ignore some useful exception-handling solutions.
In the future, we will mine exception-handling knowledge from other sources (e.g., GitHub and Stack Overflow) to continuously extend our API knowledge base.
We may also leverage the LLM as a neural knowledge base~\cite{rai2023explaining} to consult it for API exception knowledge, as the LLM ``sees'' all kinds of formal and informal API documentation during pre-training.
This will further simplify our approach and mitigate the reliance on high-quality API reference documentation.

\section{Related Work}
\subsection{Large Language Model \& Prompt Engineering}
LLMs (e.g., BERT~\cite{devlin2018bert}, BART~\cite{lewis2019bart} and GPT-3~\cite{brown2020language}) have become ubiquitous in NLP field and achieved impressive performance in various tasks including question answering~\cite{arora2022ask}, content creation~\cite{schick2022peer, yang2022re3, wu2022ai}, logical reasoning~\cite{creswell2022selection, kazemi2022lambada}, software testing~\cite{lemieux2023codamosa, kang2022large, liu2022fill} and robotics~\cite{singh2022progprompt, huang2022inner}.
PEER~\cite{schick2022peer} and Re$^3$~\cite{yang2022re3} decompose content creation as recursive plan, write and revision steps, which achieve strong performance across various domains and editing tasks.
%Wu et al.~\cite{wu2022ai} leverage LLMs to rewrite peer reviews by chaining split-points, ideation, and compose-points steps. They also demonstrate LLMs for travel flashcard creation through brainstorming, ideation, and translation. 
%Singh et al.~\cite{singh2022progprompt}  perform robotics operations by robot task planning, execution and feedback. Similar LLM-based robotics can be found in~\cite{huang2022inner}
Lemieux et al.~\cite{lemieux2023codamosa} propose CODAMOSA for test case generation, which conducts search-based software testing until its coverage improvements stall, then asks LLMs (i.e., Codex~\cite{chen2021evaluating}) to provide example test cases for under-covered functions.
Besides, LLMs offer a potential solution to automated graphical user interface (GUI) testing~\cite{liu2022fill}.
The robustness and performance of LLMs depend strongly on the quality of prompt and tremendous effort has been made on the prompt engineering~\cite{gu2021ppt, liu2022p, chen2022knowprompt, liao2022ptau}.
Nashid et al.~\cite{nashid2023retrieval} present a retrieval-based prompt selection method to solve test assertion generation and program repair problems, which can potentially be applied to multilingual and multitask settings without task or language-specific training.
%Liu et al.~\cite{liu2023pre} classify prompts according to the shape of prompts (cloze prompts), answer engineering (answered prompts) and task-specific prompts in order to clarify the paradigm of prompt-based learning.

%Prompt design is also crucial to our work.
%We introduce domain-specific knowledge from API document to prompt, which makes LLMs understand the exceptions that should be considered and finally generate more robust code. 

\subsection{Code Generation}
Following the success of LLMs at natural language tasks, the application of LLMs to code has generated significant interest.
As a result, multiple models for generating code were developed, such as Codex~\cite{chen2021evaluating}, GitHub Copilot~\cite{copilot}, AlphaCode~\cite{li2022competition}, PolyCoder~\cite{xu2022systematic}, InCoder~\cite{fried2022incoder} and CodeGen~\cite{nijkamp2022conversational}.
Codex~\cite{chen2021evaluating} is a state-of-the-art code generation model that utilizes GPT-3 technology and an API interface for public access.
GitHub Copilot~\cite{copilot} is powered by Codex and trained on public GitHub repositories, which support multiple programming languages such as Python, JavaScript, TypeScript, Ruby, and Go.
Although LLMs' utility is apparent, the degree of their robustness remains uncertain, which leads to a wide range of research interests.
Barke et al.~\cite{barke2023grounded} study how programmers interact with Copilot and provide recommendations for improving the usability of future AI programming assistants.
%Mastropaolo et al.~\cite{mastropaolo2023robustness} investigate whether using different, yet semantically equivalent, natural language descriptions would produce similar suggested functions by asking Copilot to generate Java methods based on the original Javadoc description.
Chen et al.~\cite{chen2022codet} leverages one LLM to automatically generate code samples and test cases for them.
%This work considers two factors: the consistency of the results with the generated test cases and the agreement of the outcomes with other code samples.
Recently, ChatGPT~\cite{chatgpt} has garnered a significant amount of attention, and researchers try to leverage this powerful technique to generate high-quality code~\cite{dong2023self}.
In this work, we focus on exception handling in the process of code generation to make the generated code samples more reliable. 

\subsection{Exception Handling}
Exception handling practice has been widely studied in literature~\cite{sena2016understanding, barbosa2018global, de2018studying, nguyen2019recommending, ren2020api}.
Padua et al.~\cite{de2018studying} conduct an empirical study to explore the relationship between exception handling practices and software quality on open-source Java and C\# projects.
They find that exception flow characteristics in Java projects have a significant relationship with post-release defects.
%Barbosa et al.~\cite{barbosa2018global} propose a heuristic approach that takes into account the broader context of exceptions and generates recommendations on the violations of exception handling.
Nguyen et al.~\cite{nguyen2019recommending} design a tool to predict the potential exception type that could occur in a given code snippet and recommend proper code to handle those exceptions.
To implement automated exception handling, Zhang et al.~\cite{zhang2020learning} propose a novel neural approach to predict the locations of try blocks and automatically generate the complete catch blocks for an exception.
Recent work~\cite{leinonen2023using} tends to apply LLMs to handle exceptions with explanations of the errors and suggestions on how to fix the error, and shows promising performance in this field. 
Similarly, Our work takes advantage of LLMs' few short learning abilities and obtains domain-specific knowledge from API documents to guide the exception code generation.

\section{Conclusion}
This paper presents the first knowledge-driven prompt chaining based code generation approach. 
We first extract exception-handling specifications from official API documentation and construct an API knowledge base.
Then, we use the fine-grained knowledge to construct knowledge-driven prompt chains to assist LLMs in considering exception-hand in code generation tasks.
Our \tool has been proven to be highly efficient and effective in handling exceptions.
The usefulness of our \tool-based approach has also been demonstrated in practice.
However, we acknowledge that the capability of LLMs to analyze code and the availability of exception-handling specifications in API documentation are still limited. Thus, there is still much room for improvement in this field. In the future, we aspire to address these limitations and make significant contributions to the field of knowledge-driven code generation for handling exceptions.
\section{Acknowledgements}
We would like to thank the reviewers for their detailed comments and constructive suggestions.
This research was partially supported by ``the Fundamental Research Funds for the Central Universities''(226-2022-00064).
\bibliographystyle{IEEEtran}
\bibliography{reference}
\balance
\end{document}